\documentclass{PoS}

\usepackage{amsmath,color,graphicx,multicol,braket,pdfpages,float,xcolor}
\usepackage{amssymb,latexsym,graphics,caption,pdflscape,lscape,titletoc}
\usepackage{cite,wrapfig,multirow,feynmf,trsym,booktabs,multicol,slashed}

\renewcommand{\d}{\text{d}}
\renewcommand{\t}[1]{\text{#1}}
\newcommand{\mc}[1]{\mathcal{#1}}
\newcommand{\mrm}[1]{\ensuremath{\mathrm{#1}}}

\newcommand{\saj}{s_{aj}}
\newcommand{\sjb}{s_{jb}}
\newcommand{\sab}{s_{ab}}
\newcommand{\sAB}{s_{AB}}
\newcommand{\sjk}{s_{jk}}
\newcommand{\sak}{s_{ak}}
\newcommand{\sAK}{s_{AK}}

\newcommand{\tsc}[1]{\textsc{#1}}

\newcommand{\vc}{\tsc{Vincia}}
\newcommand{\py}{\tsc{Pythia}}
\newcommand{\mg}{\tsc{MadGraph}}

\newcommand{\eqRef}[1]{eq.~\eqref{#1}}
\newcommand{\figRef}[1]{fig.~\ref{#1}}

\graphicspath{{images/}}

\title{Coherent Showers for the LHC}

\ShortTitle{Coherent Showers for the LHC}

\author{\speaker{Nadine Fischer}\\
        Monash University\\
        E-mail: \email{nadine.fischer@monash.edu}}

\author{Peter Skands\\
        Monash University\\
        E-mail: \email{peter.skands@monash.edu}}

\abstract{We present a full-fledged antenna shower for hadron collisions in the
\vc\ framework and focus on initial state radiation.
The current version of the shower is limited to massless QCD partons and colourless 
resonances.
As a proof-of-concept for the applicability of tree-level matrix-element corrections 
a la GKS we study $Z$ boson production at the LHC, corrected up to Born + 
$\mathcal{O}(\alpha_s^2)$, and show some first results with matrix-element corrections.}

\FullConference{12th International Symposium on Radiative Corrections (Radcor 2015) 
                and LoopFest XIV (Radiative Corrections for the LHC and Future Colliders)\\
                15-19 June  2015\\
                UCLA Department of Physics \& Astronomy Los Angeles, CA, USA}

\begin{document}

\section{Introduction}

Parton-shower Monte Carlo programs, which allow to predict the full final-state kinematics, play
an essential role in probing the Standard Model and searching for hints of phenomena beyond our 
current knowledge. As the LHC experiments enter the second long phase
of data collection, the experimental precision will increase, as will
the demand for precise and fast tools to provide theory predictions.

In this talk, we will describe the most important ingredients of our QCD antenna
shower, focusing on initial-initial and initial-final configurations. We will then
briefly review the matrix-element correction procedure and give some details
for $Z$ production at the LHC. Finally, we present some first, preliminary results.

\section{Antenna Shower \label{sec:shower}}

A QCD antenna~\cite{Gustafson:1987rq,Kosower:2003bh} represents a colour-connected parton pair, 
undergoing a $2\to3$ branching process.
In other shower models, such as DGLAP~\cite{Gribov:1972ri,Dokshitzer:1977sg,Altarelli:1977zs} 
and Catani-Seymour dipole~\cite{Catani:1996jh,Catani:1996vz} based ones,
one parton acts as the emitter, while one or more other partons play the role
of the recoiler(s). In the antenna picture there is no such distinction; the
antenna is treated as a single entity with a single radiation kernel. This kernel
consists of antenna functions used to drive the radiation in the shower, which can
be calculated due to the factorization of amplitudes in the soft and collinear limits.
For our choice of antenna functions see Ref.~\cite{Ritzmann:2012ca}. Since the 
factorization of amplitudes holds in the soft limit, colour coherence is an intrinsic property 
of the antenna functions.

The kinematics map or recoil strategy specifies how the two on-shell parent parton momenta
are replaced with the three on-shell daughter-parton momenta while conserving energy and 
momentum. The $(n+1)$-particle phase-space factorizes into the
$n$-particle and antenna phase-space, 
$\d \Phi_{n+1}=\d \Phi_\t{ant}~\d \Phi_{n}$, depending on the specific choice of the kinematics map.
The antenna phase-space constitutes an important ingredient in the construction of the shower.

\paragraph{Notation}
The partons will be labelled as follows: capital letters for pre-branching 
(parent) and lower-case letters for post-branching (daughter) partons,
as in $IK\to ijk$. The
first letters of the alphabet, $a$, $b$, are used for incoming partons, while
outgoing ones are denoted by $i$, $j$, $k$. We denote the recoiler
(more generally the recoiling system) by $R$ and $r$ respectively.

We restrict ourselves to massless partons and denote the branching invariant between
two partons $x$ and $y$ by $s_{xy}~\equiv~2p_xp_y~=~(p_x+p_y)^2$, which is always 
positive regardless of whether the partons involved are in the initial or final state.

We will denote the evolution variable by $t$; it is evaluated on the post-branching 
branching partons, hence, e.g., $t_{\mrm{IF}}=t(\saj,\sjk)$.

\paragraph{No-emission Probability}
A shower algorithm is based on the probability that no branching occurs between two 
scales $t_\t{start}$ and $t_\t{end}$, with $t_\t{start}>t_\t{end}$. In the case of an initial-initial
antenna $AB$ with emission $j$, the no-emission probability reads
\begin{align}
  \Pi(t_\t{start},t_\t{end})~=~
  \exp\left(-\int_{t_\t{end}}^{t_\t{start}} \d\Phi_\t{ant}~
  4\pi\alpha_s~\mc C~\frac{f_a(x_a,t)}{f_A(x_A,t)}\frac{f_b(x_b,t)}{f_B(x_B,t)}~
  \bar a\,_{AB\,j}(\saj,\sjb,\sAB)\right)~, 
  \label{eq:NoEmiProb}
\end{align}
with the strong coupling $\alpha_s$, colour factor $\mc C$, colour- and coupling-stripped antenna 
function $\bar a\,_{AB\,j}$ and the (double) ratio of PDFs,
\begin{align}
  R_\t{pdf}~=~\frac{f_a(x_a,t)}{f_A(x_A,t)}\frac{f_b(x_b,t)}{f_B(x_B,t)}~.
\end{align}

\begin{figure}[tbp]
\centering
\includegraphics[width=0.98\textwidth]{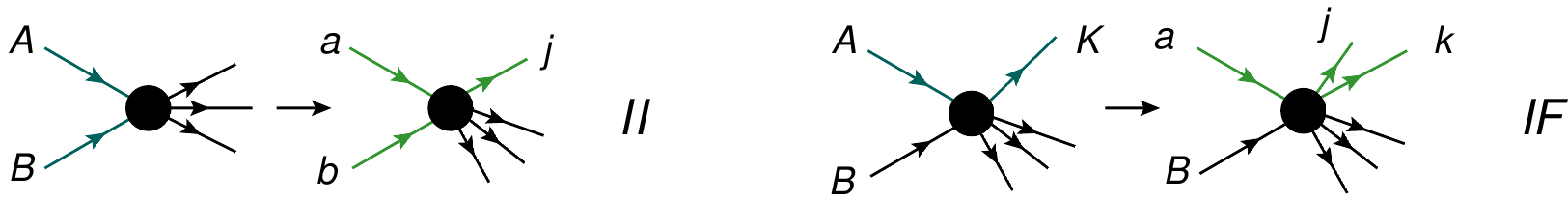}
\caption{\label{fig:notation} Illustration of initial-initial, $AB\to abj$, and initial-final branchings,
$AK\to akj$. For the II case, the recoil absorbed by the hard system is illustrated by the change in
orientation of the three outgoing lines representing the original final-state system.}
\end{figure}

\paragraph{Initial-Initial Configurations}
An initial-initial branching is denoted by $AB\to abj$ with the
recoiling system $R\to r$,
which consists of all partons produced in the collision $A+B \to R$, see the left side 
of \figRef{fig:notation}.
We define our kinematics maps to keep the direction of the beam fixed
and to preserve the invariant mass as well as the rapidity of the
recoiling system: $m_r^2 = m_R^2$ and $y_r = y_R$. 

The antenna phase-space expressed in the branching invariants reads
\begin{align}
  \d\Phi_\t{ant}^\t{II} = \frac1{16\pi^2}\,\frac{\sAB}{\sab^2}\,
  \d\saj\,\d\sjb\,\frac{\d\phi}{2\pi}~.
\end{align}

\begin{figure}[tbp]
\centering
\begin{tabular}{ll}
{\it a)} Initial-Initial Phase Space &
{\it b)} Initial-Final Phase Space \\
\includegraphics[width=0.35\textwidth]{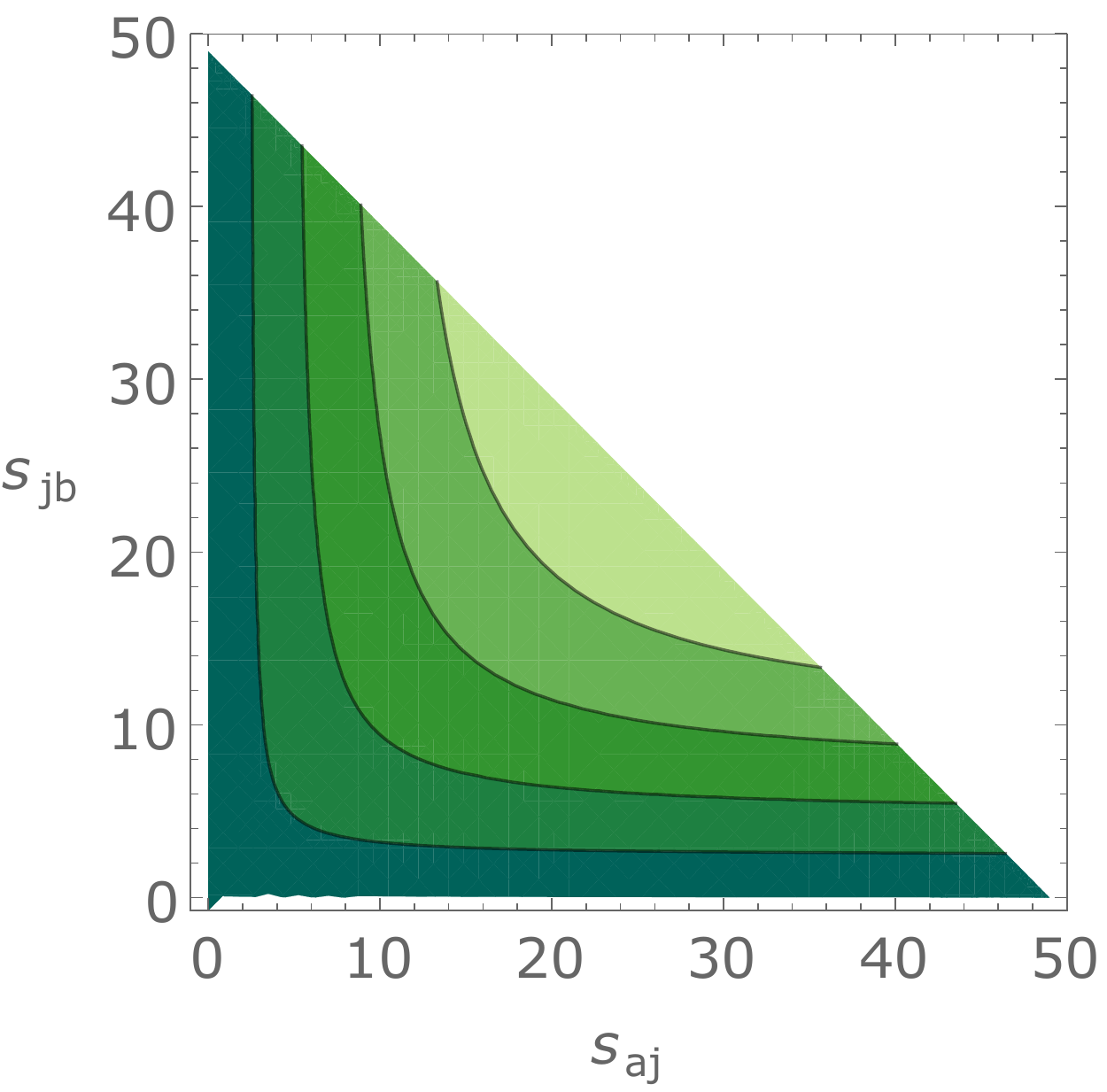} &
\includegraphics[width=0.35\textwidth]{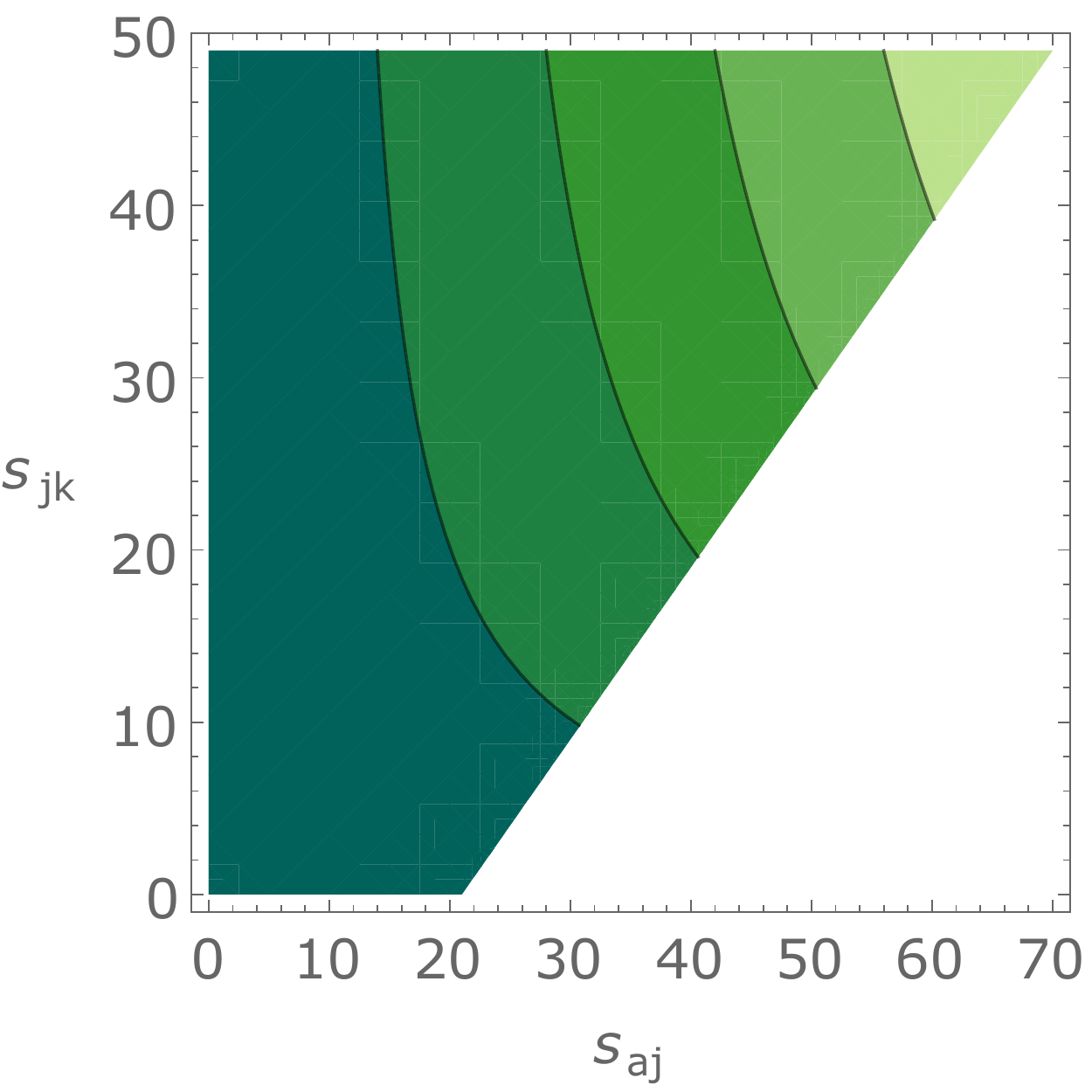}
\end{tabular}
\caption{\label{fig:phasespace} Contours of constant gluon-emission evolution
variable for {\it a)} initial-initial and {\it b)} initial-final configurations.
For {\it a)} the recoiler is chosen to be a Higgs boson, $\sAB=m_H^2$, and for
{\it b)} $\sAK=10500\,\t{GeV}^2$ and $x_A=0.3$. For both cases, the total hadronic  $\sqrt s=7\,\t{TeV}$.}
\end{figure}

The evolution variable for gluon emission is the physical transverse momentum of the emission,
\begin{align}
  t_\t{II}^\t{emit} = p_{\perp\,\t{II}}^2=\frac{\saj\sjb}\sab 
  =\frac{\saj\sjb}{\sAB+\saj+\sjb} ~,
\end{align}
which exhibits the same symmetry as the leading poles of the corresponding antenna functions. 
Constant contours of $p_{\perp\,\t{II}}^2$ are shown in \figRef{fig:phasespace}\,{\it a)}, as 
a function of the two branching invariants $\saj$ and $\sjb$. The upper phase-space bound is
$\sAB+\saj+\sjb\le s$, where $s$ denotes the hadronic centre-of-mass energy squared, and
defines the hypotenuse of the triangular shape.

For branchings with flavour changes in the initial state the antenna functions contain only 
single poles. We then use the corresponding invariant, $\saj$ or $\sjb$ respectively, as evolution
variable,
\begin{align}
  t_\t{II}^\t{conv} = Q^2_\t{II} = \left\{ \begin{array}{cl}
  \saj & \mbox{for $a$ converting to/from a gluon}\\
  \sjb & \mbox{for $b$ converting to/from a gluon}
  \end{array}
  \right.~.
\end{align}

The complementary phase-space variable $\zeta$, which has to be linearly independent of $t$ with
only one possible mapping between $(\saj,\sjb)$ and $(t,\zeta)$, is chosen to make the integrands and 
phase-space boundaries in the no-emission probability as simple and efficient as possible. An example
for gluon emission is $\zeta=\saj/\sab$.

\paragraph{Initial-Final Configurations}
We denote the pre- and post-branching partons participating in an initial-final
branching by $AK\to akj$. We define our kinematics map to keep the
direction of the beam fixed and to conserve momentum locally within
the antenna, leaving the momentum of the
spectators unchanged, $p_b=p_B$ and $p_r=p_R$, see the right-hand side
of \figRef{fig:notation}. 

The antenna phase-space expressed in the branching invariants reads
\begin{align}
  \d\Phi_\t{ant}^\t{IF} = \frac1{16\pi^2}\,\frac{\sAK}{(\sAK+\sjk)^2}\,
  \d\saj\,\d\sjk\,\frac{\d\phi}{2\pi}~.
\end{align}

We evolve gluon emission in the transverse momentum of the emission, defined as
\begin{align}
  t_\t{IF}^\t{emit} = p_{\perp\,\t{IF}}^2=\frac{\saj\sjk}{\sAK+\sjk}
  =\frac{\saj\sjk}{\sak+\saj} ~.
\end{align}
Constant contours of $p_{\perp\,\t{IF}}^2$ are shown in \figRef{fig:phasespace}\,{\it b)},
as a function of the two branching invariants $\saj$ and $\sjk$. Note that the phase-space 
is limited by $\sjk\le\sAK(1-x_A)/x_A$ and $\saj\le\sAK+\sjk$.

For branchings with flavour changes in the initial or final state we use the corresponding
invariant, $\saj$ or $\sjk$ respectively, 
\begin{align}
  t_\t{IF}^\t{conv} = Q^2_\t{IF} = \left\{ \begin{array}{cl}
  \saj & \mbox{for $a$ converting to/from a gluon}\\
  \sjk & \mbox{for $K\to q\bar{q}$} 
  \end{array} 
  \right.~.
\end{align}

As for initial-initial configurations, the complementary phase-space variable $\zeta$ is chosen 
to simplify the no-emission probability. An example for gluon emission is $\zeta=\sAK/(\sAK+\sjk)$.

\section{Matrix-Element Corrections for $pp\to Z+X$ \label{sec:MECs}}

We review the GKS procedure for iterative matrix-element corrections (MECs),
which has successfully been used to include MECs through $\mc O(\alpha_s^4)$ for hadronic $Z$ decays
in~\cite{Giele:2011cb}. Here, we apply it to initial state radiation in $pp\to Z$ up
to $\mc O(\alpha_s^2)$. 

MECs take the all-orders approximation of the shower as their starting point, 
and apply finite multiplicative correction factors to this structure order by order in perturbation 
theory. This correction factor essentially replaces the antenna
function by the corresponding leading order (LO) matrix element,
\begin{align}
  \label{eq:PnME}
  \mc O(\hat t_i,t_i)~\mc C_i\,\bar a_i~~\rightarrow~~\mc O(\hat t_i,t_i)~\mc C_i\,\bar a_i\,P_n^\t{ME}
  \qquad\t{with}\quad P_n^\t{ME}=\frac{|\mc M_n|^2}{\sum_j \mc O(\hat t_j,t_j)~\mc C_j\,\bar a_j\,|\mc M_{n-1}|^2}~,
\end{align}
with the $n$-parton matrix-element correction factor $P_n^\t{ME}$,
scale of the current branching $t_i$ and the function $\mc O(\hat
t_i,t_i)$ expressing the ordering of the shower, with respect to a
reference scale $\hat t_i$.  
The sum in the denominator of the correction factor runs over all possible ways the shower could have 
produced the $n$ parton out of the $n-1$ parton state.
Traditional showers have dead zones, as they are strongly ordered with
$\mc O(\hat t_i,t_i)=\mc O(t_{i-1},t_i)=\Theta(t_{i-1}-t_i)$,
where $t_{i-1}$, the scale of the last branching, acts as reference scale. This requires the combination 
of different event samples to fill all of phase-space. To avoid dead zones to begin with, we allow our 
shower to start at the phase-space maximum for the first branching and, starting from the second branching,
to produce unordered branchings, suppressed with the factor
\begin{align}
  \mc O(\hat t_i,t_i)=\frac{\hat t_i}{\hat t_i+t_i}~,
\end{align}
where the scale $\hat t_i$ is calculated based solely on the $n-1$ parton state at hand. $\hat t_i$
is the smallest of all scales, associated with the branchings from any
$(n-2)$ to the $(n-1)$-parton state. 
All antennae involved in the branching whose scale entered in the
determination of $\hat t_i$ are allowed
to restart their evolution at the phase-space limit, whereas all other
antennae use $\hat t_i$ as restart scale. This Markovian setup ensures that in the denominator of the MEC factor in \eqRef{eq:PnME}
the clustering has to performed only one step back, and not all the
way to the corresponding Born phase-space point. Note that, however,
the factorization scale used to evaluate the PDFs is never allowed to
become larger during the shower evolution. 

\paragraph{Shower Starting Scale and PDFs}
Note that the following paragraph represents a snapshot of what was
done during the development of the shower at the time the talk was
given. The method described here is now not our  
preferred choice, which will be explored in Ref.~\cite{VinciaNew}.
As mentioned before, to apply the MECs in $Z$ production, we start the 
shower from the phase-space maximum, the hadronic centre-of-mass
energy squared $s$. To do so  
we use the same scale as factorization scale in the hard process and apply a reweighting procedure,
resulting in the following Born exclusive cross section at the shower cutoff scale $\mu$,
\begin{align}
  \Pi(s,\mu)~f_0(x_0,m_Z^2)~\left|\mc M_Z\right|^2~\d\Phi_Z~.
\end{align}
For all higher orders we choose the reweighting such that the PDFs appear with a 
factorization scale that is the maximum of the $Z$ mass squared and the scale of the first
branching, e.g., for the Born+1 exclusive cross section with the matrix element
correction factor already applied,
\begin{align}
  \Pi(t_2^\t{max},\mu)~\frac{f_1(x_1,t_1)}{f_0(x_0,t_1)}~\Pi(s,t_1)~f_0(x_0,\t{max}(m_Z^2,t_1))
  ~\left|\mc M_{Z+1}\right|^2~\d\Phi_{Z+1}~,
\end{align}
where $t_2^\t{max}$ is the phase-space maximum for the second branching, as the shower is
allowed to produced unordered branchings.

\paragraph{An Example}
We consider the matrix-element correction to $Z$ production with gluon
emission only. We use \tsc{Rambo}~\cite{Kleiss:1985gy} to generate
large samples of uniformly distributed $pp\to Zgg$ phase-space points
and cluster them back to the corresponding $pp\to Z$  
phase-space point, using the exact inverse of the $2\to3$ recoil
prescription of the shower as a clustering algorithm. To compare the
shower approximation with the LO matrix element for 
$q_1\bar q_2\to Zg_3g_4$, we use the tree-level PS-to-ME ratio
(the inverse of the matrix-element correction factor above)
\begin{align}
  \label{eq:R3}
  R_3 =&~ \frac{\mc O(t_{\,\widehat{43}},t_3)\,\mc C_{qg}\,\bar a_{qg\,g}^\t{IF}(1,4,3)\,
  \mc C_{q\bar q}\,\bar a_{q\bar q\,g}^\t{II}(\widehat{13},2,\widehat{43})\,\left|\mc M_Z(Z)\right|^2}
  {\left|\mc M_{Zgg}(1,2;Z,3,4)\right|^2} \nonumber \\
  &~+~ \frac{\mc O(t_{\,\widehat{34}},t_4)\,\mc C_{\bar qg}\,\bar a_{\bar qg\,g}^\t{IF}(2,3,4)\,
  \mc C_{q\bar q}\,\bar a_{q\bar q\,g}^\t{II}(1,\widehat{24},\widehat{34})\,\left|\mc M_Z(Z)\right|^2}
  {\left|\mc M_{Zgg}(1,2;Z,3,4)\right|^2} \vphantom{\frac{\dfrac12}{1}}~,
\end{align}
where hatted variables denote clustered momenta. The two terms correspond to the two possible
shower paths, where either gluon 3 or gluon 4 is clustered first. The corresponding sequential
clustering scales are 
\begin{align}
  t_3 = p_{\perp\,\t{IF}}^2(g_3)~~,~~~
  t_{\,\widehat{43}} = p_{\perp\,\t{II}}^2(g_{\,\widehat{43}})
  \qquad\t{and}\qquad
  t_4 = p_{\perp\,\t{IF}}^2(g_4)~~,~~~
  t_{\,\widehat{34}} = p_{\perp\,\t{II}}^2(g_{\,\widehat{34}})~.
\end{align}

\begin{figure}[tbp]
\centering
\includegraphics[width=.496\textwidth]{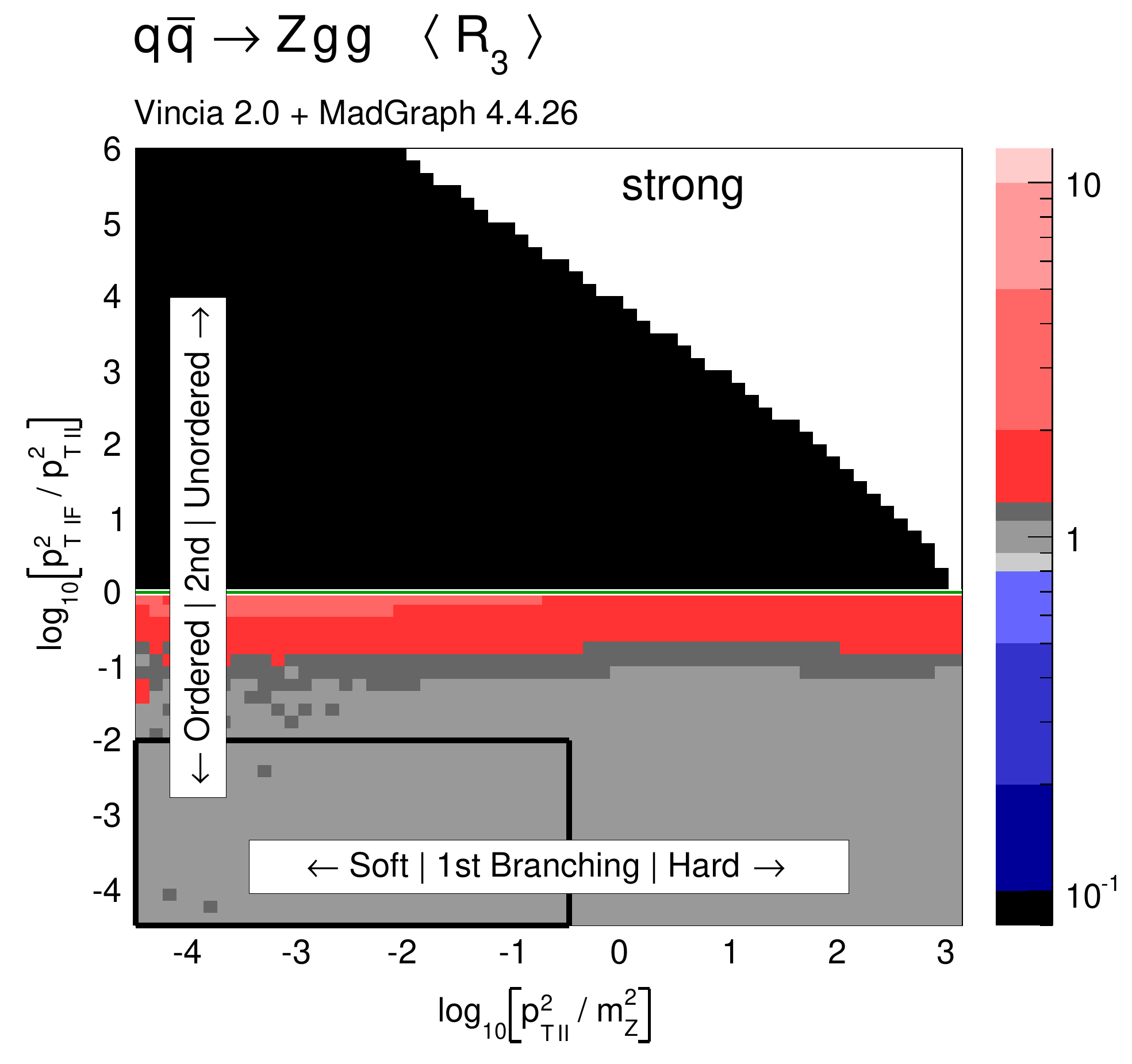}
\includegraphics[width=.496\textwidth]{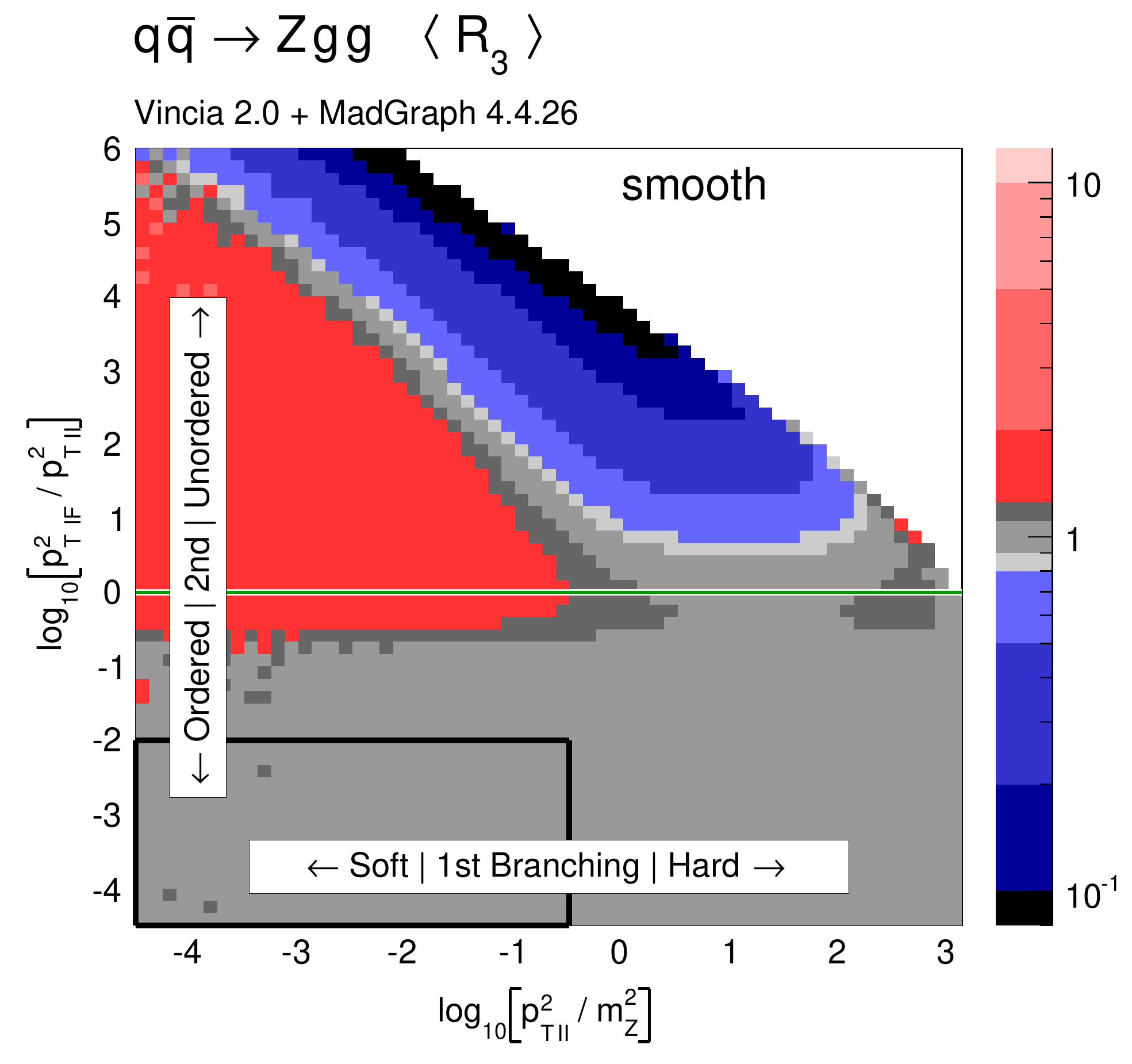}
\caption{\label{fig:PSME2D} The value of $\langle R_3 \rangle$ for $pp\to Zgg$, differentially 
over the 4-parton phase-space, with $p_\perp^2$ ratios characterizing the first and 
second emissions on the $x$- and $y$ axis, respectively. Strong (left) and smooth 
(right) ordering in the shower, with gluon emission only. Leading colour, no sum over 
colour permutations. Matrix elements generated with \mg~4~\cite{Alwall:2007st}.}
\end{figure}

In \figRef{fig:PSME2D} we show the average of the tree-level PS-to-ME ratio $\langle R_3 \rangle$ 
for $pp\to Zgg$, differentially over the 4-parton phase-space. The $x$ axis characterizes the
scale of the first emission, $p_{\perp\,\t{II}}^2$, normalized to the $Z$ mass, while the $y$
axis is the scale of the second emission, $p_{\perp\,\t{IF}}^2$, normalized to the scale
of the first emission, $p_{\perp\,\t{II}}^2$. Note that from the two possible shower paths, the more 
singular one, the one where the scale of the second emission is smaller, is used to characterize the 
phase-space point. 
The left plot in \figRef{fig:PSME2D} shows the shower with strong ordering conditions,
\begin{align}
  \mc O(t_{\,\widehat{43}},t_3) = \Theta(t_{\,\widehat{43}}-t_3)
  \qquad\t{and}\qquad
  \mc O(t_{\,\widehat{34}},t_4) = \Theta(t_{\,\widehat{34}}-t_4)~.
\end{align}
The strongly ordered shower produces a dead zone as there are no phase-space points contributing 
to $y$-values larger than zero, since those correspond to scales $p_{\perp\,\t{IF}}^2 > p^2_{\perp\,\t{II}}$.
In the strongly ordered region defined by $p_{\perp\,\t{IF}}^2 \ll p^2_{\perp\,\t{II}} \ll m^2_Z$
(the black box in the plots), the shower describes the tree-level matrix element very well, 
as expected.
The right plot in \figRef{fig:PSME2D} shows the shower with smooth ordering conditions,
\begin{align}
  \mc O(t_{\,\widehat{43}},t_3) = \frac{t_{\,\widehat{43}}}{t_{\,\widehat{43}}+t_3} 
  \qquad\t{and}\qquad
  \mc O(t_{\,\widehat{34}},t_4) = \frac{t_{\,\widehat{34}}}{t_{\,\widehat{34}}+t_4}~.
\end{align}
These conditions remove the dead zone, while not changing the shower in the strongly ordered region.
The shower is quite well behaved in the region $p_{\perp\,\t{IF}}^2 > p^2_{\perp\,\t{II}}$, but
shows some larger deviations to the tree-level matrix element at the edge of phase-space in the upper
right part of the plot. This part of phase-space corresponds to events where the $Z$ boson has been 
produced as an emission off the initial quarks rather than in the hard process. Thus, a QCD shower
will undercount the radiation in this region, as it does not include the corresponding singularities.

\section{Preliminary Results \label{sec:results}}

\begin{figure}[tbp]
\centering
\includegraphics[width=.45\textwidth]{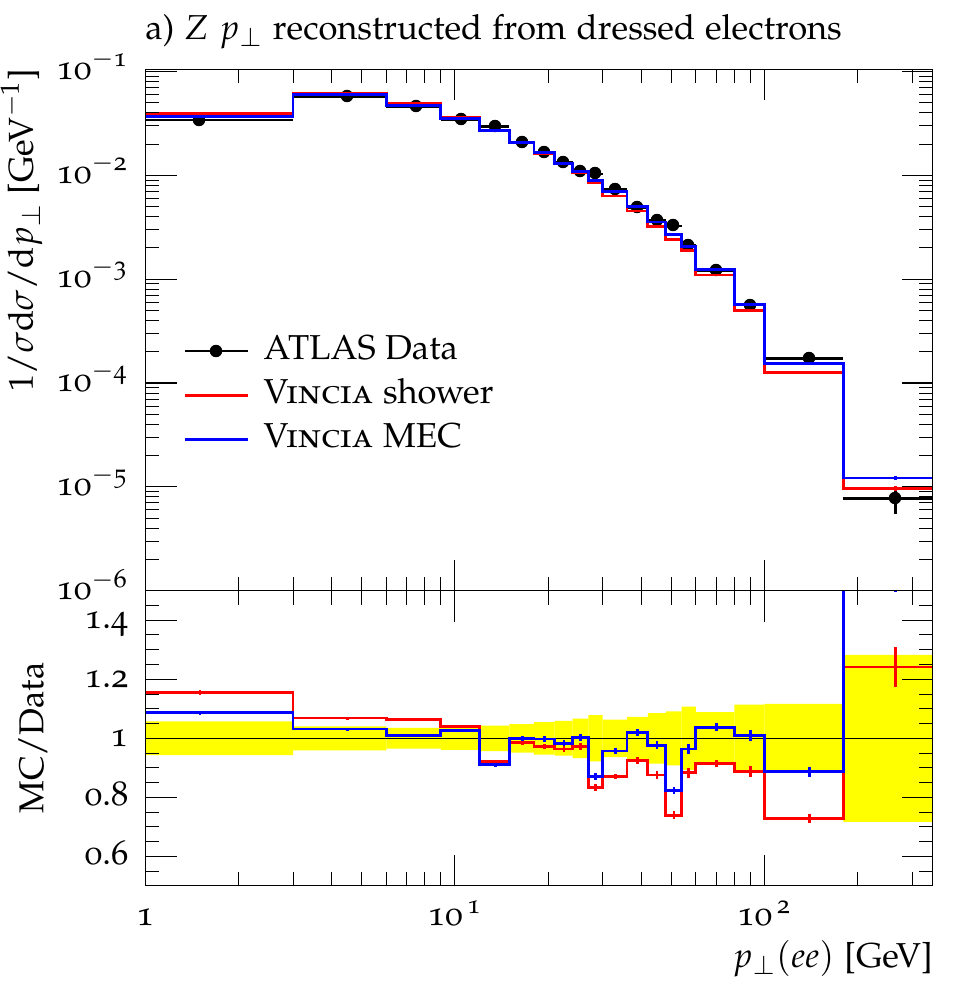} \hspace*{2mm}
\includegraphics[width=.45\textwidth]{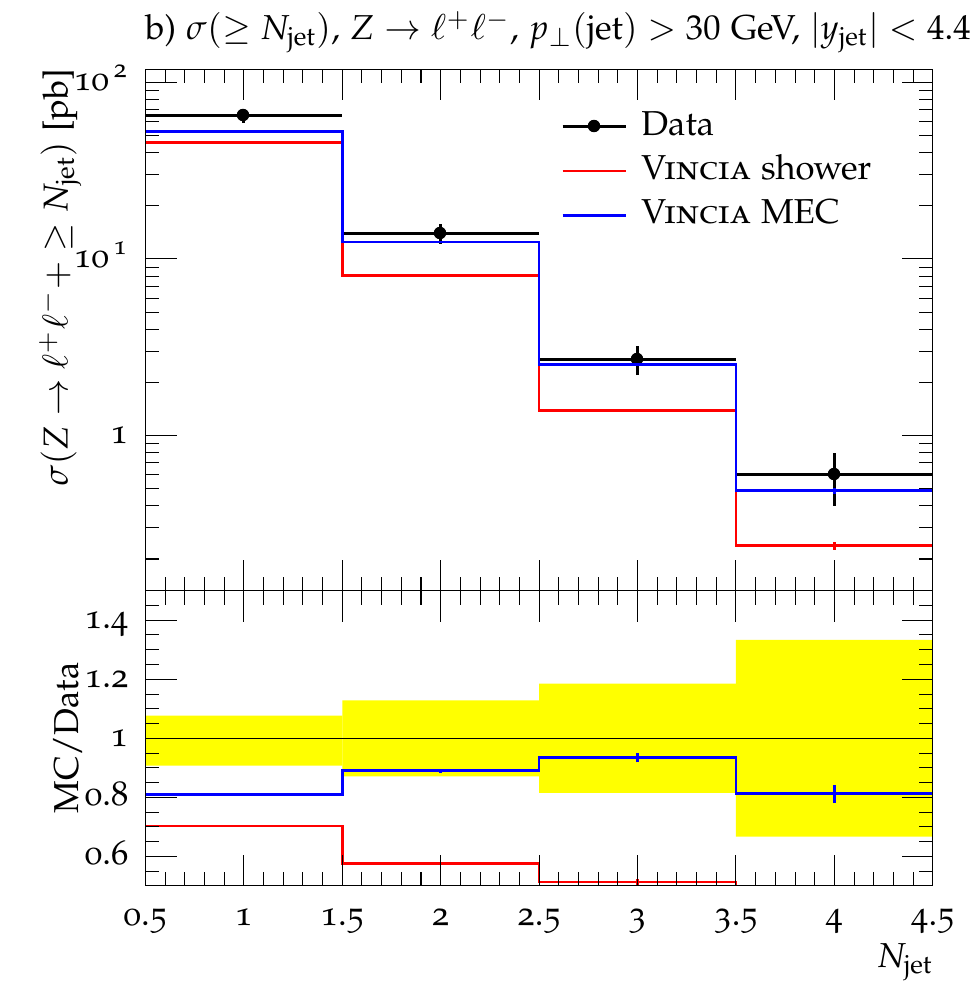}
\caption{\label{fig:res} {\it{a)}} $Z$ transverse momentum in 
$pp\to Z+X\to e^+e^-+X$ events and {\it{b)}} cross section for $pp\to Z+$~jets events.
Pure shower in red and with MECs in blue.
Analysis from \tsc{Rivet}~\cite{Buckley:2010ar}.}
\end{figure}

As some preliminary results, we show the transverse momentum of the $Z$ boson in 
$pp\to Z+X\to e^+e^-+X$ events in \figRef{fig:res}~{\it{a)}} and the cross section
for $pp\to Z+$~jets events in \figRef{fig:res}~{\it{b)}}. We use events generated by 
\py~8~\cite{Sjostrand:2014zea} and shower them with \vc. In the shower we use a one-loop 
running of $\alpha_s$ with a rather larger reference value of $\alpha_s(m_Z)=0.138$. 
For this value we find a fairly good agreement between the pure shower (red curve) and the data
for the $Z$ transverse momentum. By switching MECs on, we find an 
improved description. To get a reasonable result for the cross section involving additional
jets, we need to include the higher-order tree-level matrix elements, as indicated by 
\figRef{fig:res}~{\it{b)}}.

\section{Summary and Outlook \label{sec:outlook}}

We presented a full-fledged QCD antenna shower for initial state radiation. 
To improve the description of the shower, we use the iterated matrix-element correction procedure with the example process $pp\to Z$, corrected up to Born + 
$\mathcal{O}(\alpha_s^2)$.

The development of a more highly automated interface to
\mg~5~\cite{Alwall:2014hca}, as well as the handling of mass effects
and resonance decays, are among the main future development targets.

\acknowledgments

PS is supported in part by the Australian Research Council, contract 
FT130100744 ``Virtual Colliders: high-accuracy models for high energy physics''.

\end{document}